# Looking Back, Moving Forward: A First-Person Perspective Of How Past Artificial Intelligence Encounters Shape Today's Creative Practice

**Makayla Lewis**
Digital Media Kingston, Kingston University
London, United Kingdom
m.m.lewis@Kingston.ac.uk

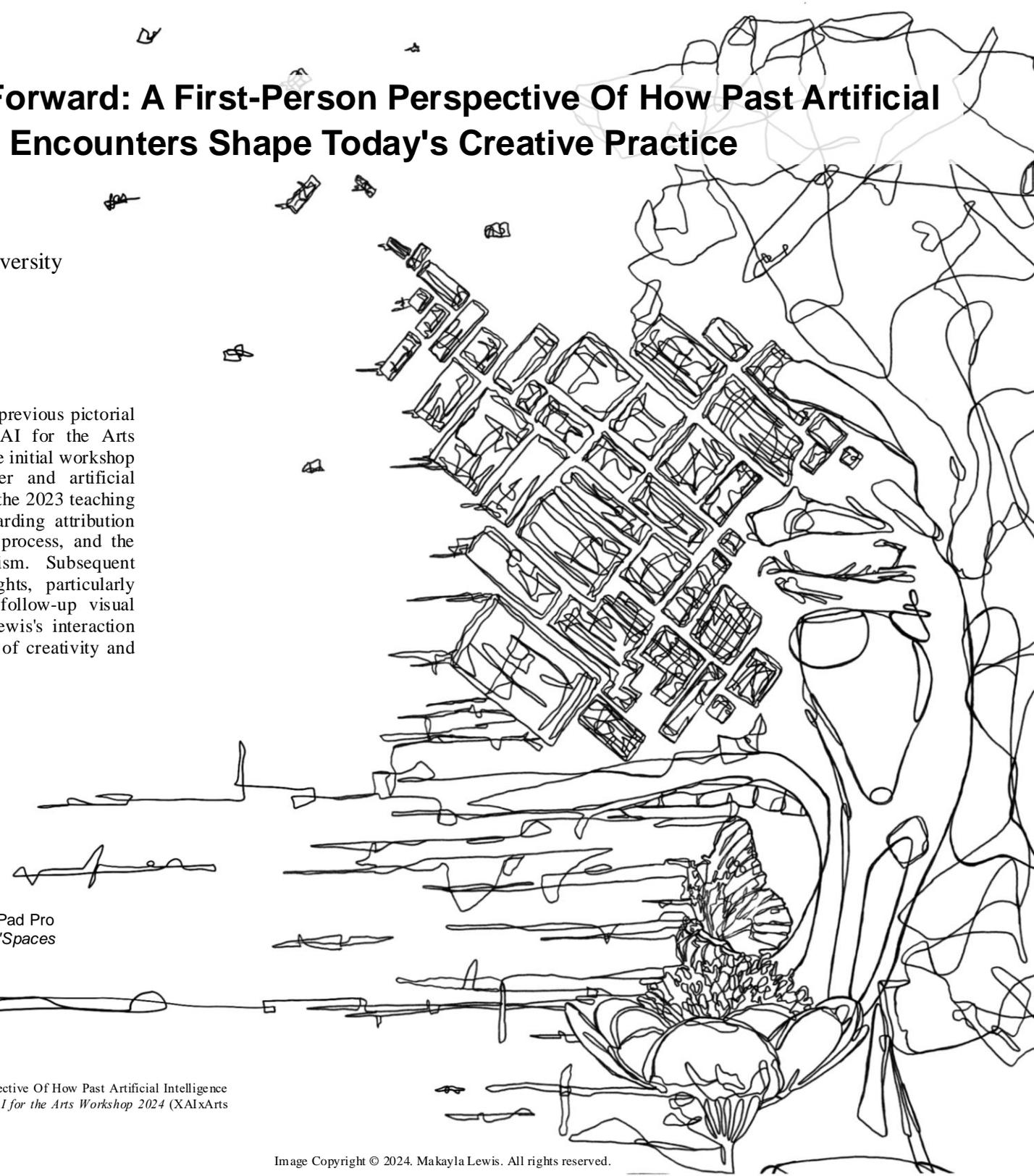

**ABSTRACT**
This visual narrative is a first-person reflection of the previous pictorial at the 1st International Workshop on Explainable AI for the Arts (XAIxArts) at ACM Creativity and Cognition 2023. The initial workshop pictorial explored a relationship between researcher and artificial intelligence, navigating creative challenges throughout the 2023 teaching block. It concluded by raising crucial questions regarding attribution transparency, the ethical dimensions of the creative process, and the delicate balance between inspiration and plagiarism. Subsequent discussions at the workshop yielded valuable insights, particularly concerning interpreting the creative journey. This follow-up visual narrative reflects the enduring impact of Makayla Lewis's interaction with AI. A self-portrait that delves into the interplay of creativity and introspection.

**CSS Concepts**
• Human-centered computing ~ Human computer interaction (HCI) ~ Empirical studies in HCI.

**Figure 1:** Looking Back, Moving Forward. Apple Pencil on iPad Pro 11 using Procreate by Makayla Lewis, 2024; exhibited at *"Spaces of Enquiry"* in Stanley Picker Gallery, United Kingdom [16].

## INTRODUCTION

Artistic practice has been an enduring cornerstone of human civilisation, serving as a conduit for expressing and interpreting sensory experiences across various mediums, including painting, music, performance, sculpture, etc. Its multifaceted nature allows for the communication, documentation, and personal expression of our experiences, transcending cultural and physical boundaries. Digital technology has further democratised artistic creation and facilitated global accessibility and instantaneous sharing of artwork, e.g. [2, 3, 4, 6]. However, the creative process is susceptible to challenges, particularly when daily experiences intersect with artistic output, giving rise to phenomena like "art block" or "creative block," which hinder the creative flow [14]. In the context of the *1st International Workshop on Explainable Artificial Intelligence for the Arts (XAIxArts)* at ACM Creativity and Cognition 2023, Makayla Lewis presented a personal narrative titled 'AIxArtist: A First-Person Tale of Interacting with Artificial Intelligence to Escape Creative Block' [11]. This narrative recounted their experience interacting with artificial intelligence, specifically *Chat GPT* and *MidJourney*, to overcome a period of creative stagnation induced by the demands of a hectic teaching term.

### First - Person Reflection

This interaction with artificial intelligence reignited Makayla Lewis's first-person, e.g., [5, 9, 10, 11, 12, 17, 18] creative process, providing guidance and inspiration akin to an art teacher offering prompts and feedback to students [11]. Reflecting on this experience a year later, the author observed a shift in their artistic trajectory. While the prompts initially sparked creativity, they did not leave a lasting impact. Instead, Makayla Lewis found themselves drawn towards introspective practices, exploring self-examination and emotional expression themes in their artwork. One particular *Chat GPT* prompt stood out - *to create a self-portrait*. This challenge inspired Makayla Lewis to delve into their internal landscape, resulting in a dynamic portrayal that encapsulated feelings of pressure and introspection. This departure from traditional observational drawing marked a pivotal moment in their artistic journey, redirecting focus towards internal exploration rather than external observation, which they have struggled to set aside.

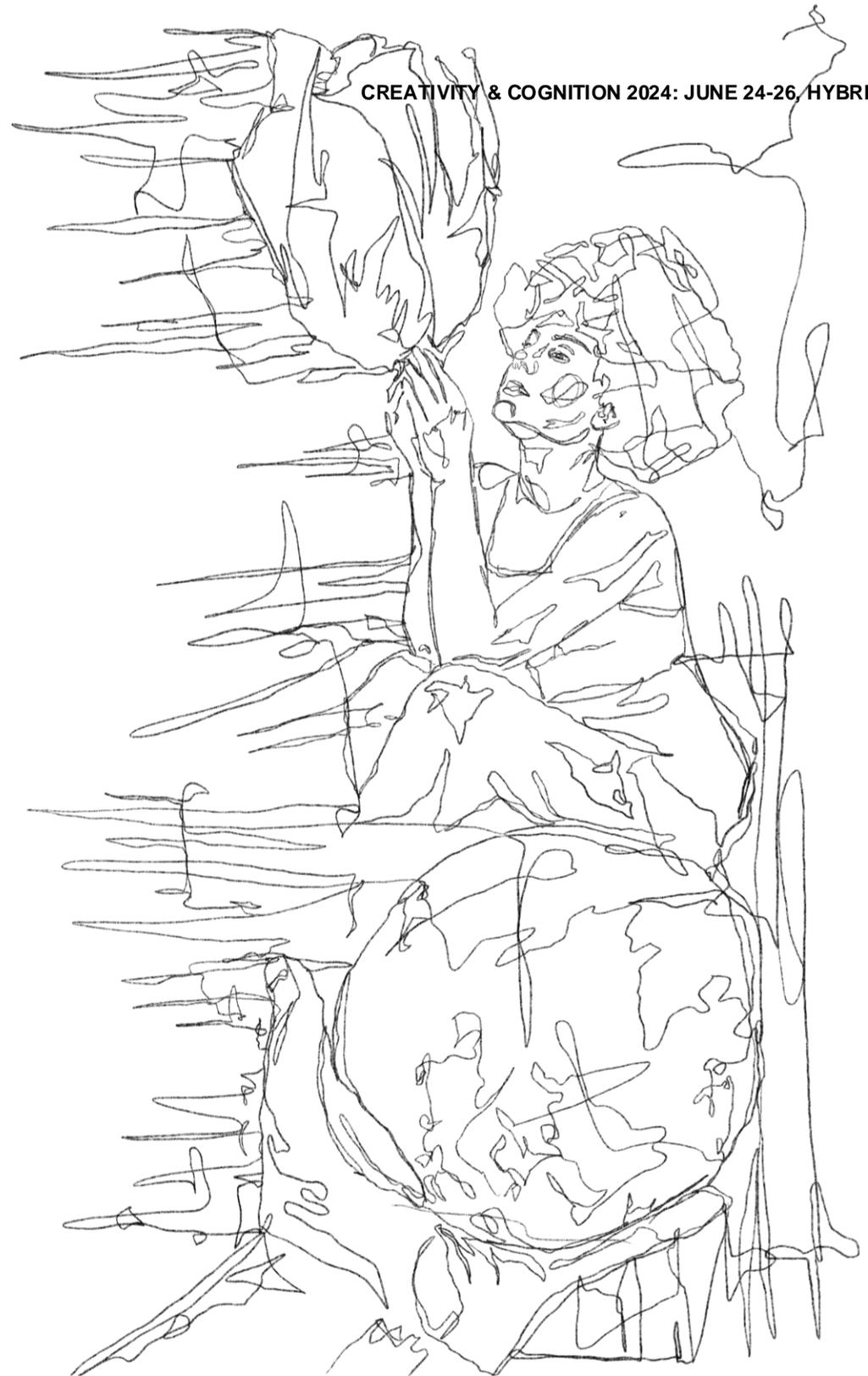

**Figure 2:** Social media and I. Digital Sketchbook Entry. Apple Pencil on iPad Pro 11 using Procreate by Makayla Lewis, 2023.

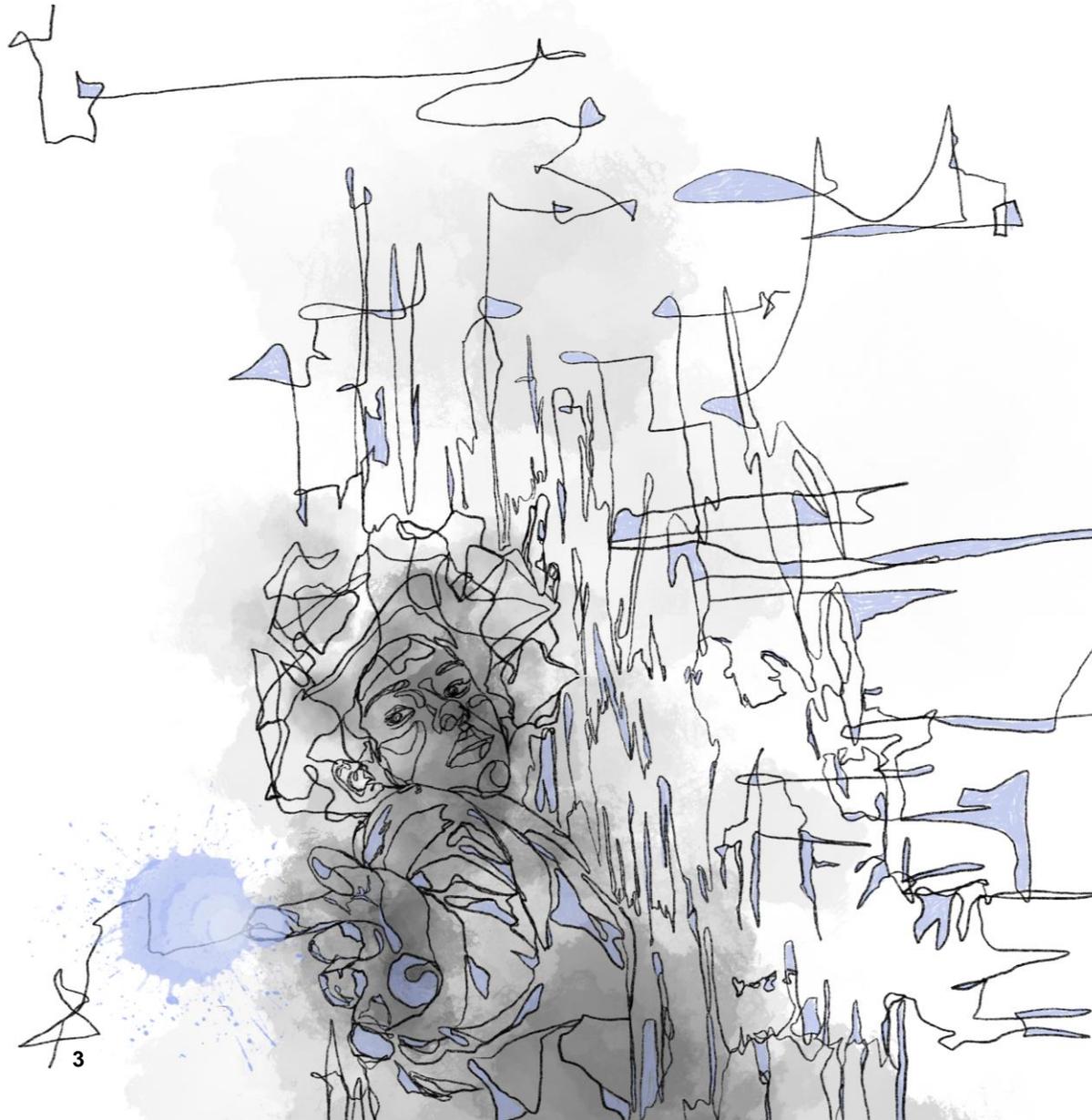

**Figure 3:** What does inclusion feel like? Commission for Kingston University London Inclusion in STEM event. Apple Pencil on iPad Pro 11 using Procreate by Makayla Lewis, 2024.

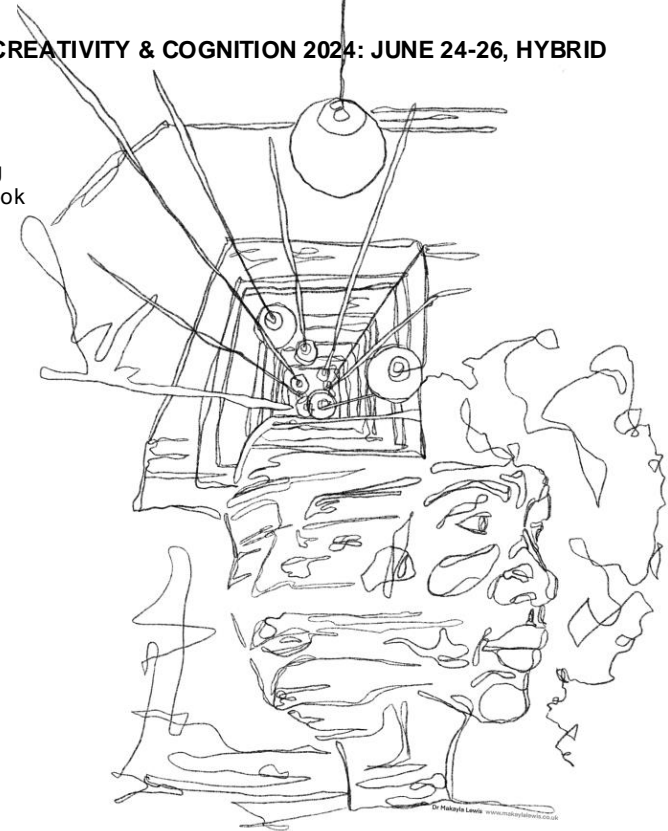

**Figure 4:** TLE. Sketchbook entry sharing what epilepsy feels like. Digital Sketchbook Entry. Apple Pencil on iPad Pro 11 using Procreate by Makayla Lewis, 2023.

### SELF-PORTRAIT

Throughout history, artists have frequently depicted others in their portraits, often for various purposes, such as capturing the likeness of prominent figures or conveying social status. For example, Leonardo da Vinci's *Mona Lisa* is a renowned portrait that captures the enigmatic smile of its subject, believed to be Lisa Gherardini, the wife of Florentine merchant Francesco del Giocondo [13]. However, there have been instances where artists have turned their brushes inward, capturing their likeness [1]. One notable example is Vincent van Gogh, who famously painted self-portraits. These self-portraits are believed to serve multiple purposes, including self-reflection, experimentation with colour and technique, and documentation of his physical and emotional state. One of the most iconic self-portraits by van Gogh depicts him with a *Bandaged Ear* [7, 19], painted in January 1889, shortly after his release from the hospital. This portrayal came in the aftermath of a desperate act, wherein van Gogh severed most of his left ear during a heated altercation with a fellow painter [7, 19]. The painting documents this tumultuous event and provides a window into van Gogh's psyche and unwavering commitment to his artistic practice [7, 19].

**INTROSPECTION & CREATIVE PRACTICE**

Upon encountering these self-portraits in person nearly a decade ago at their respective galleries, Makayla Lewis felt a visceral connection to van Gogh's determination to continue creative practice. This prompted introspection on the nature of artistic expression and communication, e.g., [15]. Is the creative block a thing, or is another more personal vehicle required to *"keep the creative juices"* going? As illustrated in this pictorial (Figures 2, 3, 4, 6, and 7), this journey of introspection through self-portraiture has prevented creative block from occurring during teaching block 2024, *"All I needed was "someone" to tell me to draw myself, it just happened to be Chat GPT ."*

**Moving forward, looking back**

As 2024 unfolded, Makayla Lewis was invited to participate in the Spaces of Enquiry gallery exhibition at Stanley Picker Gallery in the United Kingdom [16]. This exhibit, which brought together art students and scientific researchers from Kingston University London, aimed to explore collaborative ways of understanding the world and fostering new avenues of investigation. Among the works displayed was Makayla Lewis' initial pictorial piece [11], alongside works of four students responding to their and other artists' work over the years.

This context prompted a reflection on the original self-portrait: **Would a new rendition convey the same message as before?** The original piece was described as an *"Abstract self-portrait in response to ChatGPT and MidJourney"*, featuring dynamic lines merging human and robotic elements [11]. However, the subsequent realisation was unexpected: the new portrait took on an organic quality, departing from the anticipated narrative (see Figure 6 for work-in-progress sketches and Figure 1 for the final artwork exhibited at [16]).

**Figure 5:** Where do I fit - Exploring the fight to find a space for Makayla Lewis art in the structured/redefined HCI academic world. A Commission for Arts in HCI Artbook [18]. Apple Pencil on iPad Pro 11 using Procreate by Makayla Lewis, 2023.

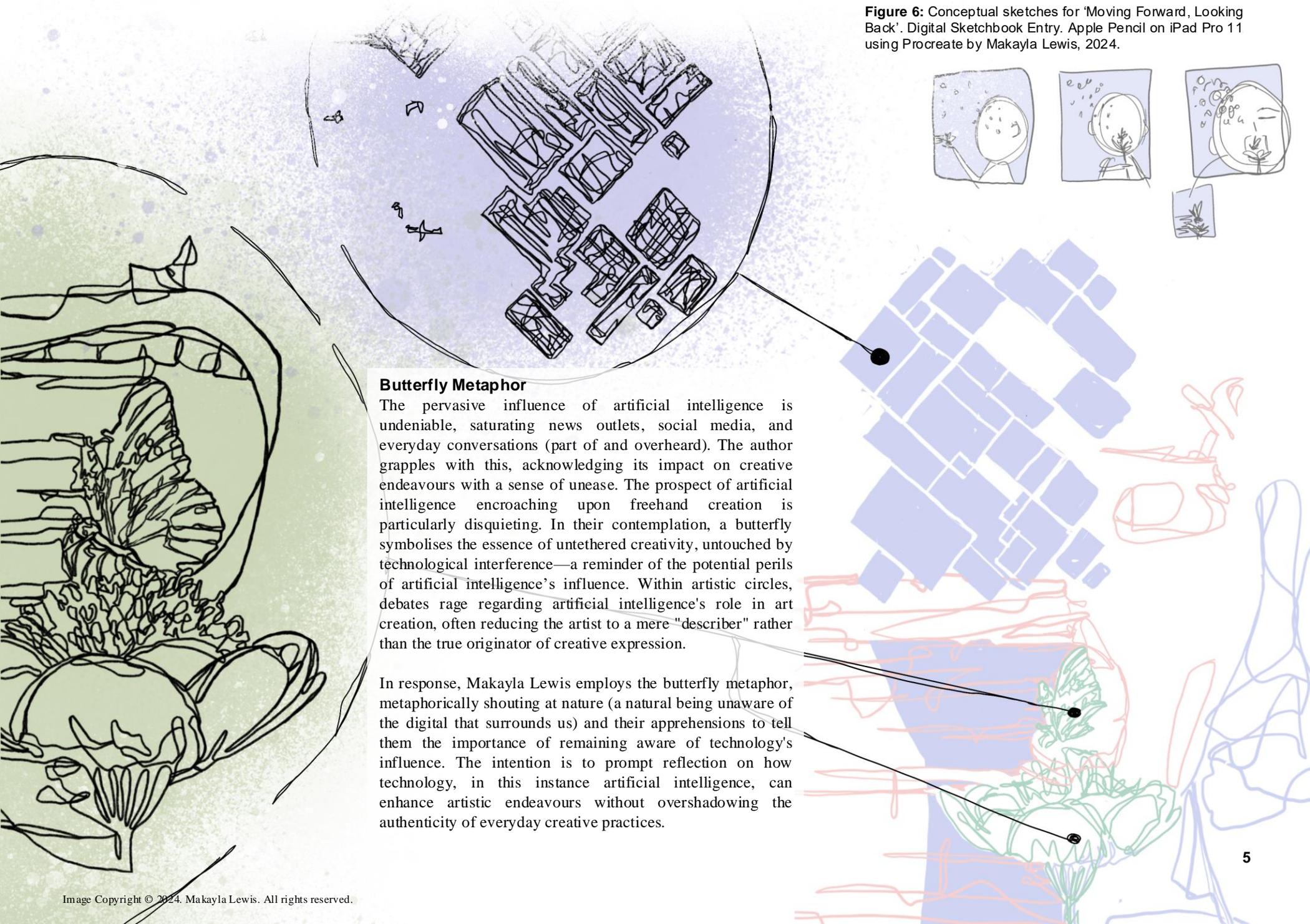

**Figure 6:** Conceptual sketches for 'Moving Forward, Looking Back'. Digital Sketchbook Entry. Apple Pencil on iPad Pro 11 using Procreate by Makayla Lewis, 2024.

**Butterfly Metaphor**

The pervasive influence of artificial intelligence is undeniable, saturating news outlets, social media, and everyday conversations (part of and overheard). The author grapples with this, acknowledging its impact on creative endeavours with a sense of unease. The prospect of artificial intelligence encroaching upon freehand creation is particularly disquieting. In their contemplation, a butterfly symbolises the essence of untethered creativity, untouched by technological interference—a reminder of the potential perils of artificial intelligence's influence. Within artistic circles, debates rage regarding artificial intelligence's role in art creation, often reducing the artist to a mere "describer" rather than the true originator of creative expression.

In response, Makayla Lewis employs the butterfly metaphor, metaphorically shouting at nature (a natural being unaware of the digital that surrounds us) and their apprehensions to tell them the importance of remaining aware of technology's influence. The intention is to prompt reflection on how technology, in this instance artificial intelligence, can enhance artistic endeavours without overshadowing the authenticity of everyday creative practices.

**CLOSING COMMENTARY**

This workshop position pictorial does not take a definitive stance on these debates. Instead, it offers insights into a recalibrated approach to creative practice—an introspective journey to preserve the integrity of artistry. While artificial intelligence-generated prompts and imagery may seem formulaic, Makayla Lewis recognises their potential to ignite innovative thought and action. The depicted artistic journey serves as a testament to the unexpected outcomes of artificial intelligence interaction, prompting introspection and sparking newfound avenues of exploration. Was this outcome by design on the part of artificial intelligence? Unlikely. Were they consciously seeking a divergent path? Not necessarily. Yet, the dialogue between human creativity and artificial intelligence yields a pleasantly surprising introspective experience that may prove enriching. Finally, the author asks, *What do you see in Figure 1?*

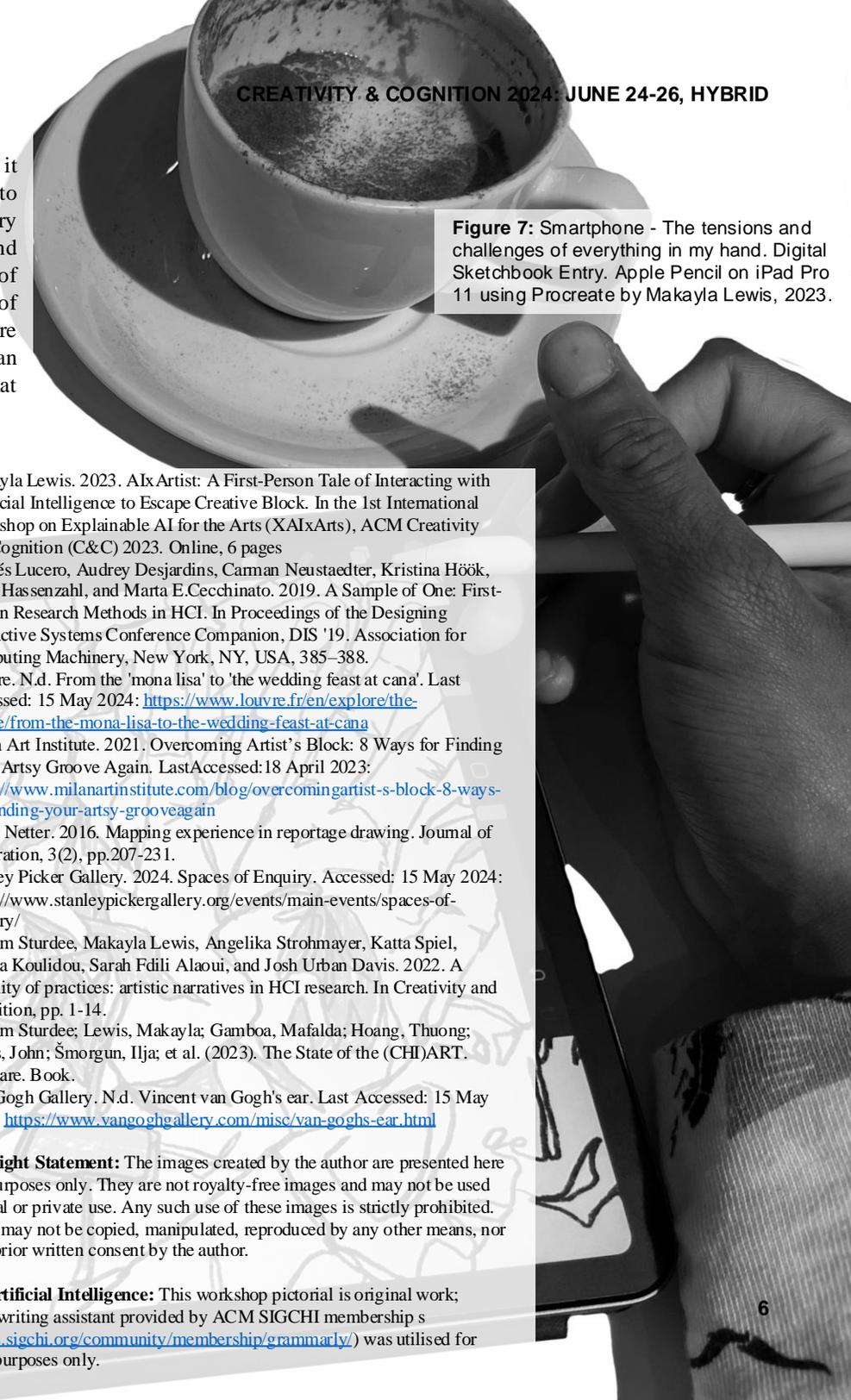

**Figure 7:** Smartphone - The tensions and challenges of everything in my hand. Digital Sketchbook Entry. Apple Pencil on iPad Pro 11 using Procreate by Makayla Lewis, 2023.

**Disclosure Artificial Intelligence:** This workshop pictorial is original work; Grammar.ly (writing assistant provided by ACM SIGCHI membership s https://archive.sigchi.org/community/membership/grammarly/) was utilised for copy-editing purposes only.